\documentclass[prd,aps,showpacs,nofootinbib,nobibnotes,superscriptaddress,preprintnumbers]{revtex4}

\usepackage{color}
\usepackage{amsmath}
\usepackage{graphicx}

\usepackage[breaklinks]{hyperref}
\usepackage[english]{babel}
\usepackage{graphicx}
\usepackage{color}
\usepackage{amsfonts,amsthm}
\usepackage{bm,bbm}
\usepackage{soul}
\usepackage{mathtools,slashed}
\usepackage{footmisc}

\newcommand{\mn}{\ensuremath{{_{\mu\nu}}}}

\newcommand{\pa}[1]{\ensuremath{\partial_{#1}}}

\newcommand{\al}{\ensuremath{\alpha}}
\newcommand{\be}{\ensuremath{\beta}}
\newcommand{\ga}{\ensuremath{\gamma}}
\newcommand{\Ga}{\ensuremath{\Gamma}}

\newcommand{\la}{\ensuremath{\lambda}}
\newcommand{\na}{\ensuremath{\nabla}}
\newcommand{\sig}{\ensuremath{\sigma}}
\newcommand{\bpsi}{\ensuremath{\bar{\psi}}}

\newcommand{\lnm}{\ensuremath{\Lambda_{Q}}}
\newcommand{\lrbg}{\ensuremath{\Lambda_{\rm RBG}}}

\newcommand{\viu}[2]{\ensuremath{{e^{#1}}_{#2}}}
\newcommand{\vid}[2]{\ensuremath{{e_{#1}}^{#2}}}

\newcommand{\lag}{\mathcal{L}}

\newcommand{\lr}[1]{\left(#1\right)}
\newcommand{\lrsq}[1]{\left[#1\right]}

\newcommand{\bee}{\begin{equation}}
\newcommand{\ee}{\end{equation}}
\newcommand{\ba}{\begin{eqnarray}}
\newcommand{\ea}{\end{eqnarray}}
\newcommand{\bal}{\begin{align}}
\newcommand{\eal}{\end{align}}

\definecolor{Darkgreen}{RGB}{30,120,30}

\def\bs{\begin{subequations}}
\def\es{\end{subequations}}

\def\la{\lambda}

\def\cO{\mathcal{O}}

\newcommand{\book}[5]{\emph{#1} (#2, #3, #4, #5)}

\newcommand{\tia}[1]{}

\newcounter{listcounter}

\begin{document}

\title{Effective interactions in Ricci-Based Gravity below the non-metricity scale}

\author{Adria Delhom}
\affiliation{Departamento de F\'{i}sica Te\'{o}rica and IFIC, Centro Mixto Universidad de
Valencia - CSIC. Universidad de Valencia, Burjassot-46100, Valencia, Spain}

\author{Victor Miralles}
\affiliation{Departamento de F\'{i}sica Te\'{o}rica and IFIC, Centro Mixto Universidad de
Valencia - CSIC. Universidad de Valencia, Burjassot-46100, Valencia, Spain}

\author{Ana Pe\~nuelas}
\affiliation{Departamento de F\'{i}sica Te\'{o}rica and IFIC, Centro Mixto Universidad de
Valencia - CSIC. Universidad de Valencia, Burjassot-46100, Valencia, Spain}

\begin{abstract}
We  show how minimally-coupled matter fields of arbitrary spin, when coupled to Ricci-Based Gravity theories, develop non-trivial effective interactions that can be treated perturbatively only below a characteristic high-energy scale $\lnm$. Our results generalize to arbitrary matter fields those recently obtained for spin 1/2 fields in \cite{Latorre:2017uve}. We then use this interactions to set bounds on the high-energy scale $\lnm$ that controls departures of Ricci-Based Gravity theories from General Relativity. Particularly, for Eddington-inspired Born-Infeld gravity we obtain the bound $ |\kappa|< 3.5\times10^{-14} \text{ m}^5 \text{kg}^{-1}\text{s}^{-2} $.
\end{abstract}
\maketitle

\section{Introduction}
The equivalence principle suggests the understanding of gravitation as geometrodynamics, an idea which has proven to be very powerful, lying at the heart of General Relativity (GR) and most of the alternative theories of gravity that have been proposed in the past century. After a century of Eddington's expedition \cite{CrispinoEddington} that served as its first confirmation, GR has performed outstandingly in predicting all the gravitational phenomena experimentally observed up to date\footnote{It would be fair to say that, though the missing mass problem and the expansion of the universe can be perfectly explained within GR, extra ingredients which have not been found yet are needed to account for this phenomena, and other solutions that imply modifications of the gravitational sector have also been taken into account.}. Despite the success of GR in explaining experimental data, there are many (theoretical) reasons to go beyond GR, such as  the information paradox, the understanding of the nature of space-time singularities, or the search for a quantum theory of gravity \cite{Hawking:1974sw,Hawking:1976ra,Mathur:2008wi,tHooft:1984kcu,OritiBook}. GR was built under the assumption that the affine connection is completely determined by the metric (i.e. the \textit{Riemannian/metricity} postulate). Such assumption was natural at the time GR was formulated, as only Riemannian (or metric) geometries were known, and implies  that all the geometric information is encoded in the metric. Shortly after the birth of GR the torsion tensor was introduced in differential geometry. Until that moment, it was thought that the only possible affine connection was the Levi-Civita connection of a metric, thus being the one employed in the construction of GR. The existence of the torsion tensor pointed out that the connection and the metric are actually two independent objects, and it gave birth to the study of non-Riemannian geometries\footnote{Non-Riemannian geometries are characterized precisely by the independence between the affine connection of a non-Riemannian manifold and its metric structure.} \cite{EisenhartNRiem}. After non-Riemannian geometries were developed, it became clear  that the assumption that the connection must be the Levi-Civita connection of the metric is an unnecessary postulate that must be experimentally tested \cite{GeometryAndExperience}. Up to date, the validity of the Riemannian postulate is experimentally well established at low-energy scales (large volumes), although it remains untested at high energy scales (small volumes) \cite{Will:2014kxa,Berti:2015itd,Daniel:2010ky,Krawczynski:2012ac,Damour:1991rd,Kramer:2006nb,Clifton:2005aj,Abbott:2016blz}. Some approaches to quantum gravity suggest that geometric structures other than the metric could be needed to account for all space-time properties at energy scales higher than the ones currently tested in gravitational experiments \cite{Olmo:2015bha,Olmo:2015wwa,Lobo:2014nwa,Hossenfelder:2017rub}.\\

In order to be able to explore the physical consequences of non-Riemannian geometrical structures, one must first understand what are the differences between Riemannian and non-Riemannian manifolds. From a mathematical point of view, such differences are encoded in two geometrical tensors that measure departures from Riemannianity: the torsion tensor ${S^\lambda{}_{[\mu\nu]}}\equiv 2{\Ga^\lambda{}_{[\mu\nu]}}$, and the non-metricity tensor $Q_{\lambda\mu\nu}\equiv -\na_{\lambda}g_{\mu\nu}$, which by definition vanish in Riemannian space-times. Once these differences are characterized within these two tensor fields, one is ready to study the role that they might play in gravitational physics. Given a gravitational action, one may or may not assume the Riemannian postulate before deriving the gravitational dynamics. This choice gives rise to two different frameworks under which one can study any theory of gravitation: If one assumes the postulate, it is said that one employs the \textit{metric} formalism, whereas if not, one employs the \textit{metric-affine} formalism. By using the metric framework any gravitational action will give rise only to Riemannian space-times and therefore, in order to explore the physics associated to non-Riemannian terms, one must employ the metric-affine formalism. In practice, while in the metric formalism one derives the dynamics of the theory from the action by first applying the metric postulate and then varying the action with respect to the metric, in the metric-affine formalism one treats both metric and connection as fundamental fields and obtains their dynamics by varying the action with respect to both of them. \\

A wide variety of modifications to GR can be characterized by higher-order curvature terms in the action. This alternative theories have been studied since the middle of the last century, with the aim of solving some of the theoretical problems listed above, and more recently for giving an explanation to the dark phenomena as well, without a clear success in their resolution (see \cite{Lobo:2008sg,Clifton:2011jh,Olmo:2011aa,Cembranos:2015svp,Joyce:2016vqv,Nojiri:2017ncd} and references within). From a physical perspective, higher-order curvature terms are well motivated for several reasons. For instance, in order to renormalize matter fields in curved space-time,  $R^{\al\be\mu\nu}R_{\al\be\mu\nu}$, $R^{\mu\nu}R_{\mu\nu}$ and $R^2$ terms must be present in the effective action \cite{Utiyama:1962sn, ParkerToms, BirrellDavies}. It has also been proven that gravity theories with quadratic curvature terms are renormalizable and asymptotically free in the metric framework \cite{Stelle:1976gc,Julve:1978xn,Fradkin:1981iu}, although they suffer from ghost-like degrees of freedom which break unitarity \cite{Antoniadis:1986tu, Johnston:1987ue}. The origin of these ghost degrees of freedom can be traced back to the fact that for higher-order curvature theories of gravity, the metric formalism gives rise to fourth-order equations of motion for the metric \cite{Woodard:2006nt,Woodard:2015zca} \footnote{With the well known exception of Lovelock theories \cite{Borunda:2008kf} \label{fn:repeat}.}. In fact, one of the original motivations to study gravitational theories within the metric-affine framework is that, even for higher-order curvature Lagrangians, the field equations for the metric are always second order, which could in principle alleviate the ghost problem, although it has been recently proved that this is not aways the case \cite{BeltranJimenez:2019acz}. In the past fifteen years, the metric-affine framework has gained more attention due to some interesting results concerning the avoidance of space-time singularities in black-hole scenarios or in Big-Bang cosmologies even at a classical level and coupled to standard matter sources \cite{Barragan:2009sq,Poplawski:2010kb,Barragan:2010qb,Olmo:2012nx,Olmo:2013gqa,Odintsov:2014yaa,Olmo:2015dba,Olmo:2015bya,Olmo:2016fuc,Olmo:2017fbc,Bambi:2015zch,Menchon:2017qed,Bejarano:2017fgz}. Given the fact that some approaches to quantum gravity present similar bouncing solutions, it has been suggested that metric-affine gravities could be understood as a low-energy limit of a possible quantum theory of gravity \cite{Olmo:2015bha,Olmo:2015wwa,Lobo:2014nwa,Hossenfelder:2017rub,Olmo:2008nf}.\\

First works in gravitation which took into account non-Riemannian geometries dealt with the possibility of including the torsion tensor in the description of gravitation, and it was seen that fermions naturally generate torsion when coupled to GR \cite{Kibble:1961ba}. Afterwards, it was  discovered that one can obtain the same results by constructing the gauge theory of the Poincar\'e group\footnote{Known as Einstein-Cartan-Sciamma-Kibble or ECKS theory.} \cite{Hehl:1976kj,Hehl:1994ue}. Some authors tried then to understand what could be the observable consequences of torsion and used them to place experimental constraints to the possible existence of a non-vanishing torsion tensor in different contexts \cite{Shapiro:2001rz,Mao:2006bb,March:2011sa,Lehnert:2013jsa,Lucchesi:2015mga,Iorio:2015rla,Boos:2016cey}. Despite the effort put in understanding the observable effects of torsion, existence (or lack) of observables related to non-metricity is not yet well understood. The first works that included non-metricity, only took into account a special case of it, namely the Weyl vector ($Q_{\al\mu\nu}=2A_\al g\mn$) \cite{Weyl}. Later on, non-metricity was studied as a gauge potential arising in the gauge theory of the group of affine transformations \cite{Hehl:1976kj,Hehl:1994ue}. More recently, modifications of the GR Lagrangian have been studied using the metric-affine formalism, where non-metricity is not constrained to vanish. In this regard, although the metric-affine version of GR has field equations identical to those of the metric version \footnote{If one considers coupling spinors to GR this is not strictly true, since there arise torsion corrections sourced by spin-density.}  \cite{Borunda:2008kf,Bernal:2016lhq,Janssen:2019htx} and non-metricity vanishes in both frameworks, higher-order curvature theories generally have non-trivial non-metricity tensors\footref{fn:repeat}, and it is still not clear what its physical implications might be. It was recently pointed out that in a broad class of theories of gravity named Ricci-based gravity theories (RBG), non-metricity corrections perturbatively induce effective interactions in spin $1/2$ fields, which were used to experimentally constrain the allowed parameters of the Lagrangians of the RBG family \cite{Latorre:2017uve}. {This class of gravitational models is defined by a Lagrangian that is an arbitrary analytic function of the metric and the Ricci tensor which features diffheomorphism and projective symmetries; and it encompasses a wide variety of models that are currently under active research; such as Eddington-inspired-Born-Infeld gravity models (see \cite{BeltranJimenez:2017doy} for a review and \cite{Bouhmadi-Lopez:2017lbx,Menchon:2017qed,Jana:2017ost,deHaro:2017yll,Escamilla-Rivera:2018soo,Shaikh:2018yku,Shaikh:2018cul,Ping-Li:2018mrt,Jana:2018knq,Avelino:2019esh,Delhom:2019zrb,Albarran:2019ssh,Nascimento:2019qor} for recent results), Palatini $f(R)$ and $f(R, R^{(\mu\nu)}R_{(\mu\nu)})$ theories (see \cite{Olmo:2011uz} for a review and \cite{Iosifidis:2018zjj,Olmo:2019qsj,Antoniadis:2019jnz,Toniato:2019rrd,Vignolo:2019qcg,Olmo:2019flu,Gialamas:2019nly,Bombacigno:2019nua,Coumbe:2019fht,Feola:2019zqg,Olmo:2015bya,Olmo:2016fuc,Olmo:2015dba,Olmo:2017fbc,Bambi:2015zch} for recent results), Weyl gravity (see \cite{Jizba:2019oaf,Maeder:2020phk,Sanomiya:2020svg,Jawad:2020wlg,Zou:2020psp} for recent results) or others. Generic properties of the full class of RBG theories have also been recently studied in \cite{BeltranJimenez:2017doy,Jimenez:2015caa,Afonso:2017bxr,Afonso:2018hyj,Afonso:2018mxn,Afonso:2019fzv,Delhom:2019btt}, finding the remarkable result that they can generically be mapped into GR with a modified matter sector. Indeed, it has been shown that the modifications that introduce RBGs with respect to GR do not change the number of degrees of freedom of the theory, but rather they modify the way in which the matter sector interacts. In this sense, note that introducing other irreducible pieces of the Riemann, such as {\it e.g.} the Weyl tensor or the antisymmetric Ricci tensor will generally excite new gravitational degrees of freedom. Thus we see that the modifications introduced by the (symmetrized) Ricci tensor are of different nature than those introduced by other covariant objects associated to the affine connection, and must be studied in detail separately.} Contact interactions induced by gravity {in Palatini $f(R)$} have already been studied by Flanagan \cite{Flanagan:2003rb} within the scalar-tensor representation of $1/R$, though these results were questioned by Vollick\footnote{{The main argument by Vollick is that field re-definitions may not always be allowed in curved space-times, however, the proof of the} {equivalence theorem for the S-matrix \cite{Kamefuchi:1961sb,Chisholm:1961tha} that ensures that observables are invariant under field redefinitions does not seem to rely on Minkowski space-time, and no argument was provided in the discusion on why this could be a crucial point. In any case, the equivalence theorem should hold in particle physics experiments, where curvature effects can be neglected to a high degree of accuracy}.} \cite{Vollick:2003ic,Vollick:2004ws}. Nonetheless, due to the fact that non-metricity in $f(R)$ is of the Weyl kind and that $f(R)$ theories are projectively invariant, these contact interactions cannot be interpreted as induced by non-metricity in the context of $1/R$ where non-metricity can be gauged away by means of a projective transformation \cite{Olmo:2011uz,Afonso:2017bxr,Iosifidis:2018zjj}. This is not the case for RBG theories more general than $f(R)$ models, which generally feature non-metricity which cannot be gauged-away by a projective transformation. Experimental constraints related to the non-metricity tensor have also been recently considered in the context of Lorentz symmetry breaking. However it was assumed that the non-metricity tensor has constant non-vanishing vacuum expectation value which breaks Lorentz invariance.  This assumption allowed  to set bounds on the existence of a background non-metricity tensor \cite{Foster:2016uui}. Given that the non-metricity of RBG models exactly vanishes in vacuum, the conclusions from such work do not apply in this context. However, notice that for any gravity theory whose dynamics shows a Lorentz violating non-metricity by means of a vacuum expectation value or otherwise, then the conclusions of \cite{Foster:2016uui} would apply to such theory\footnote{{For instance, a first approach to the metric-affine formulation of bumblebee gravity, which could potentially feature a non-vanishing vev for the non-metricity tensor, was considered in \cite{Delhom:2019wcm}.}}\\

The purpose of this paper is to show that the modifications of the space-time metric induced by non-metricity-related terms in Ricci-based theories of gravity can be understood, below the non-metricity energy scale $\lnm$, as effective operators for fields of every spin which give rise to new contact interactions, as well as using these operators to set observational bounds on the RBG family. The conclusions presented here generalize the results obtained in \cite{Latorre:2017uve} for spin $1/2$ fields. The structure of the paper is as follows. In Sec. \ref{sec:RBG} we will introduce RBG theories and then show how non-metricity appears in this context. In Sec. \ref{sec:NMLag} we will show how in a $(1/\lnm)^{4n}$ expansion of the space-time metric the $n>0$ terms, which are 
directly related to the non-metricity tensor, consist of effective operators of dimension $4(n+1)$. Since they appear through the metric, these operators couple the matter stress-energy tensor to (at least) the kinetic term of all matter fields. As particular examples, we will explicitly derive the effective operators that generate self-interactions for scalar and vector fields, and interactions between two fermions and two vectors. In Sec. \ref{sec:ExpCons}, we will couple the Standard Model (SM) to a generic RBG, and we will use the corresponding effective Lagrangians to compute the lowest-order contributions of these corrections to the scattering processes $e^{-}\ga \to e^{-}\ga$ and $\ga\ga\to\ga\ga$. These results will allow us to set experimental constraints to the RBG parameters, which directly translate into lower bounds to the scale $\lnm$ once a specific RBG theory is chosen. We will then use our bounds for the RBG family to constrain a particular RBG model which has recently attracted much interest, namely Eddington-Inspired Born-Infeld gravity (EiBI).\\
  
\section{Ricci-Based Gravities and the non-metricity tensor}\label{sec:RBG}
Let us begin with a short discussion of the field equations of RBG theories and the particular form that the non-metricity tensor adopts within this class of gravity models. The action of (projectively-invariant) RBGs is given by
\begin{equation}\label{RBGaction}
S=\frac{1}{2\kappa^2}\int d^4 x\sqrt{-g} F_{\rm RBG} \left[ g\mn,R_{(\mu\nu)}, \lrbg \right] +S_m \left[ g_{\mu\nu},\Psi,{\Gamma^\al{}_{\mu\nu}} \right]\ ,
\end{equation}
where $F_{\rm RBG}$ is any analytic scalar function of $g_{\mu\nu}$ and  $R_{(\mu\nu)}$,  $\kappa=M^{-1}_{{\rm P}}$ is the Einstein constant (with $M_{\rm P}$ the Planck mass), and $\lrbg$ is a (high-)energy scale which characterizes deviations from GR. It is useful to define the non-metricity scale $\lnm=\sqrt{P_{\rm P}\Lambda_{\rm RBG}}$ and it is the scale at which the non-metricity tensor (and thus deviations from GR) become non-perturbative in RBG models. Here $R_{(\mu\nu)}$ is defined as a function of the connection $\Ga^\al{}_{\mu\nu}$, which is a priori independent of the metric {in the metric-affine (or Palatini) approach}. The reason why {we only consider the symmetric part of the Ricci tensor in the action (which amount to consider only the projectively-invariant sub-class of the Ricci-based family) is because, as shown recently in \cite{BeltranJimenez:2019acz,BeltranJimenez:2020sqf}, the inclusion of its antisymmetric part generally unleashes the propagation of ghostly degrees of freedom related to the projective mode of the connection, and therefore the non-projectively invariant theories within the RBG family are not healthy from a physical point of view\footnote{ Let us comment that it was shown in \cite{Buchdahl:1979ut}, and later re-analised in \cite{Vitagliano:2010pq}, that the torsion-free version of the model $R+\alpha R_{[\mu\nu]}R^{[\mu\nu]}$ can be written as GR with an extra dynamical Proca field. This result was generalized in \cite{BeltranJimenez:2019acz,BeltranJimenez:2020sqf}, where it was seen that even in the case of an RBG with broken projective symmetry, in which the antisymmetric part of the Ricci tensor is included in the action, the torsion-free constraint turns the ghosts associated to the projective mode into of a healthy Proca field.}. This is also the case for scalar-tensor metric-affine theories \cite{Aoki:2019rvi}. Let us also point out that typically RBGs are constructed in agreement with the principles of Metric Theories of Gravity (see \cite{Thorne:1970wv,Will:2014kxa}), and therefore any gravitational field other than the metric typically does not couple to the matter fields,} {which ensures that the theories satisfy Einstein's Equivalence Principle. Here we depart from the principles of Metric Theories of Gravity by allowing for a coupling between matter fields and connection, which would break the EEP. However, as we will see later, this coupling is not relevant for the physical scenarios that we are interested in.}\\

The field equations of a general RBG with minimally coupled matter fields are obtained by varying the above action \eqref{RBGaction} with respect to metric and connection. In  \cite{BeltranJimenez:2017doy} (see \cite{DelhomILatorre:2021naw} for a more general derivation), it is proven that by performing appropriated field redefinitions, and integrating out the connection field equations, (projectively-invariant) RBG theories admit an Einstein frame representation, where the field equations are
\begin{align}
&{G^\mu}_\nu(q)=\frac{\kappa^2}{|\Omega|^{1/2}}\left[{T^\mu}_\nu-\delta^\mu_\nu\left(\frac{F_{\rm RBG}}{2\kappa^2}+\frac{T}{2}\right)\right]\; , \label{EinsteinlikeEqs}\\
&\nabla_{\al}\left[\sqrt{-q} q^{\mu\nu}\right]{-\delta^{\mu}{}_{\al} \nabla_{\rho}\left[\sqrt{-q} q^{\nu \rho}\right]=\kappa^2\Delta_{\al}{}^{\mu \nu}+\sqrt{-q}\left[S^{\mu} {}_{\lambda \alpha}q^{\nu \alpha}+S^{\alpha}{}_{\alpha \lambda} q^{\nu \mu}-\delta^{\mu} {}_{\lambda}S^{\alpha}{}_{\alpha \beta} q^{\nu \beta}\right]}   \label{Connectioneq}.
\end{align}

Here ${T^\mu}_\nu\equiv g^{\mu\alpha}T_{\alpha\nu}$, $T={T^\mu}_\mu$, and ${G^\mu}_\nu(q)\equiv q^{\mu\alpha}G_{\alpha\nu}(q)$; where $q_{\mu\nu}$ is an auxiliary metric defined by the relation $\sqrt{-|q|}q^{\mu\nu}\equiv \sqrt{-g}{\partial F_{\rm RBG}}/{\partial R_{\mu\nu}}$, and $|\Omega|\equiv |q|/|g|$. {Also $2\Delta_{\al}{}^{\mu\nu}\equiv  \delta S_m/\delta\Gamma^{\al}{}_{\mu\nu} $ is the hypermomentum current which accounts for the coupling between the matter fields and the affine connection} see {\it e.g.} \cite{HehlHypMom} . Once one replaces $q_{\mu\nu}$ by $g_{\mu\nu}$, the second equation is identical to the one satisfied by the connection in {metric-affine GR when coupled to matter with arbitrary hypermomentum.} Thus the connection must be the Levi-Civita connection of $q_{\mu\nu}$ (up to a projective mode) \cite{Afonso:2017bxr,Borunda:2008kf} {plus corrections that involve the hypermomentum current (see e.g. \cite{BeltranJimenez:2020sqf,Vitagliano:2010sr,Iosifidis:2018jwu})}. {For minimally coupled bosonic fields \cite{Delhom:2020hkb}, the hypermomentum vanishes, and therefore the solution to \eqref{Connectioneq} has vanishing torsion, but carries non-metricity since $\nabla_\mu g_{\al\be}$ is non-vanishing. For minimally coupled fermionic fields the non-metricity will also be non-vanishing, but also the non-vanishing fermionic hypermomentu will source a totally antisymmetric torsion term of the form $S_{\al\mu\nu}=-i\kappa^2|\Omega|^{-1/2}\epsilon_{\al\mu\nu\sigma}\lrsq{\bar{\Psi}\gamma^\sigma\gamma_5\Psi}$. As we will explain later, the effects of this torsion term induced by the fermionic hypermomentum will be of sub-leading order in the physical scenario considered in this work, and thherefore we will neglect this term in what follows.}\\

 From the definition of $q_{\mu\nu}$ and the RBG field equations, it can be shown  that it is always possible to find an on-shell relation between the space-time metric and the auxiliary metric of the form $g_{\mu\nu}=q_{\mu\al}{(\Omega^{-1})^\al}_\nu$ \cite{BeltranJimenez:2017doy}. This new matrix relating the two metrics is called the \textit{deformation matrix}, and it is completely specified once a specific RBG Lagrangian is chosen. It is crucial to note that the deformation matrix is an on-shell function of the stress-energy tensor which always admits a $1/\lnm$ expansion of the form  \begin{equation}\label{OmegaSeries}
{(\Omega^{-1})^\al}_\nu={\delta^\al}_\nu+\frac{1}{\lnm^4}\lr{\al T{\delta^\al}_\nu+\be{T^\al}_\nu}+\mathcal{O} (\lnm^{-8}),
\end{equation}
where the first term in the expansion must be $\delta^{\alpha}_{\nu}$ if we want to recover GR as a low-energy limit of the corresponding RBG model. In fact, from \eqref{EinsteinlikeEqs} it is clear that all RBG have exactly the same dynamics as GR in vacuum at a classical level, given that their field equations become identical when $T^\mu{}_\nu=0$. In the above expansion, the different RBG models are characterized by the value of the dimensionless coefficients $\alpha$ and $\be$, which are completely specified once a particular RBG Lagrangian is chosen. As required for consistency, it can be seen that $\al$, $\be$ and all higher-order coefficients in \eqref{OmegaSeries} exactly vanish if the GR Lagrangian $F_{\rm RBG}=g^{\mu\nu}R_{\mu\nu}$ is chosen.\\

Using the definition of non-metricity and \eqref{OmegaSeries}, we can see that within RBG models the non-metricity tensor takes the form $Q_{\lambda\mu\nu}=-\na_{\lambda}\lr{q_{\mu\al}{{\Omega^{-1}}^\al}_\nu}$. Therefore, we can use \eqref{Connectioneq} and \eqref{OmegaSeries} and write the metric and non-metricity tensors as

\begin{align}
&g_{\mu\nu} = q_{\mu\nu} +\frac{1}{\lnm^4}\bigg( \al T q_{\mu\nu}+\beta T\mn\bigg)+\cO(\lnm^{-8}), \label{metricstressenergy}\\
&Q_{\la\mu\nu} = \frac{1}{\lnm^4}\bigg(\alpha(\na_{\la}T) q\mn+\beta \na_{\la}T\mn\bigg) +\cO(\lnm^{-8})\, , \label{nonmetricitystressenergy} 
\end{align}
up to order $\Lambda_Q^{-8}$ corrections. Let us point out that, as a consequence of the equations of motion for the connection \eqref{Connectioneq} and the expansion of the deformation matrix \eqref{OmegaSeries}, the $1/\lnm^{4n}$ (with $n>0$) corrections to the metric are directly related to the non-metricity tensor. The existence of these terms implies a non-metricity of the form \eqref{nonmetricitystressenergy}, and vice versa, the existence of a non-metricity like \eqref{nonmetricitystressenergy} implies the $1/\lnm^{4n}$ corrections to the metric in \eqref{metricstressenergy}. In light of this, we can thus understand the perturbative effects of non-metricity within RBG theories by taking $1/\lnm$ as a small coupling. As a remark, notice that despite the fact that the $\alpha$-dependent term in (\ref{nonmetricitystressenergy}) comes from a vectorial non-metricity that can be gauged away from the connection by performing a projective transformation, the term with $\beta$ is a genuine contribution of non-metricity which cannot be eliminated from the connection by means of any symmetry. However, let us point out that unless conformal re-scalings of the metric are a symmetry of the specific RBG theory, the effects associated to $\alpha$ can still arise because of its presence in \eqref{metricstressenergy}. Thus, for the $\alpha$ term to be unphysical, the theory must have projective as well as conformal symmetry.\\

It is now pertinent to discuss the role of the auxiliary metric $q_{\mu\nu}$. Notice that we can write the gravity Lagrangian $F_{\rm RBG}$ as an on-shell function of the matter fields and $q\mn$, which allows
us to make an analogy between \eqref{EinsteinlikeEqs} and the Einstein equations of GR. Indeed, the GR field equations for the space-time metric $g\mn$ are $G^\mu{}_\nu(g)=\kappa^2 T^\mu{}_\nu$ are identical to the field equations for the auxiliary metric $q\mn$ in RBG theories (see \eqref{EinsteinlikeEqs}) but coupled to a non-linearly modified matter sector. As shown in \cite{BeltranJimenez:2017doy,Afonso:2018bpv,Afonso:2018hyj,Afonso:2018mxn}, the difference between GR and a generic RBG theory is that the right-hand side of \eqref{EinsteinlikeEqs} can be seen as a modified stress-energy tensor if one writes $\Omega$ and $F_{\rm RBG}$ as functions of the matter fields (which can be done on-shell). Given this parallelism between the field equations of $g_{\mu\nu}$ in GR and those of $q_{\mu\nu}$ in RBG theories, the role of the auxiliary metric $q_{\mu\nu}$ in RBG is exactly analog to that of the space-time metric $g_{\mu\nu}$ within GR. More explicitly, $q_{\mu\nu}$ is the gravitational field that can be associated to the usual long-range gravitational interactions (exchange of gravitons \cite{Jimenez:2015caa}). Once understood the physical meaning of the auxiliary metric, it becomes apparent that we can write it as $q_{\mu\nu}\approx \eta_{\mu\nu}+\delta q_{\mu\nu}$, where $\delta q_{\mu\nu}$ encodes the corresponding Newtonian and post-Newtonian corrections to Minkowski space-time in a given RBG model \cite{Olmo:2005hc,Olmo:2006zu}. Notice that the effects associated to $\delta q_{\mu\nu}$ can always be eliminated locally by a suitable choice of coordinates. On the contrary, the effects associated to the local distributions of energy and momentum induced by the deformation matrix $\Omega^{\mu}{}_{\nu}$ cannot be eliminated in this way. \\

In light of the above discussion it is clear that within RBG theories the space-time metric is associated to two different kinds of phenomena, namely, 1) the propagation of gravitons, which is responsible for the standard Newtonian and post-Newtonian effects associated to the space-time curvature generated by the integration over the matter sources, and 2) new effects associated to the local distribution of matter stress-energy, which are intimately related to the existence of a non-metricity tensor of the form \eqref{nonmetricitystressenergy}. Let us emphasize with some extra care the two roles that the space-time metric plays in Ricci-based theories of gravity. Point 1) refers to the standard effects of gravity, while point 2) is something new and characteristic of metric-affine theories such as RBGs. Gravitation is usually understood as an attractive long-range interaction mediated by a massless spin 2 field (the graviton) described by perturbations to the space-time metric $g_{\mu\nu}$ around a given background.  These effects are sourced by total amounts of mass/energy and become stronger the higher the total mass/energy of the sources. In the classical limit, they correspond to the usual Newtonian and post-Newtonian corrections described by GR \cite{Olmo:2005hc,Olmo:2006zu}. In RBG theories this long-range interaction associated to a spin 2 field, now described by perturbations of the auxiliary metric $q\mn$, is only a part of the space-time metric $g_{\mu\nu}$ as is apparent from \eqref{metricstressenergy}. The rest of the space-time metric is described by terms sensitive to the local distribution of energy-density, which are point-wise functions of the $T^\mu{}_\nu$. Moreover, from the structure of the field equations in RBGs, the existence of these corrections is necessarily related to a non-vanishing non-metricity tensor of the form \eqref{nonmetricitystressenergy}, which suggests naming them as non-metricity-induced corrections. These corrections are a novel feature that can distinguish RBG theories from other alternatives to GR, and understanding whether they also arise in more general theories is currently ongoing work.\\

As pointed out in \cite{Latorre:2017uve}, the non-metricity-induced corrections to the metric open a new window to look for new gravitational effects in the high-energy-density regime, which is not necessarily the usual strong (gravitational) field regime where the post-Newtonian corrections become dominant (see \cite{Olmo:2019qsj,Afonso:2019fzv} for astrophysical examples). Indeed, in order to look for non-metricity-induced corrections this picture suggests to look for scenarios with weak gravitational fields, where space-time curvature effects  (i.e. Newtonian and post-Newtonian terms) are negligible, but high-energy-density processes occur. These conditions are actually realized in particle accelerators on Earth's surface, since in those experiments the Newtonian and post-Newtonian terms due to the gravitational field of the Earth or of the interacting particles can be neglected, thus having $q_{\mu\nu}\approx \eta_{\mu\nu}$. Therefore, within an arbitrary RBG theory, the space-time around Earth's surface will be described by a Minkowskian background with small departures from GR described by the $1/\lnm^{4n}$ corrections in \eqref{metricstressenergy} which may become relevant for some high-energy processes.

\section{Effective interactions in RBG theories}\label{sec:NMLag}

We have already explained the equations that govern the dynamics of RBG theories, showing how the non-metricity tensor induces corrections in the metric $g\mn$ which are related to local (instead of global) energy-momentum-density. Therefore, we are now in the position of understanding the phenomenology related to this geometrical object (the non-metricity tensor) within RBG theories. In \cite{Latorre:2017uve} it was found that these corrections can be understood as perturbative effective interactions below the scale $\lnm$ for spin 1/2 fields. Here we will generalize this result, showing how these non-metricity-induced effective interactions are not specific of spin 1/2 fields, but a rather general consequence associated with the non-metricity tensor within RBG theories. In the following section we will show how effective interactions arise for scalar, fermion and vector fields by deriving the corresponding effective Lagrangians.  After these Lagrangians are derived, it will be clear that they induce effective interactions between any two pairs particle-antiparticle appearing in an arbitrary matter sector. This interactions are such that they respect all the symmetries of the original matter action.  To give some examples, we will derive the operators contributing to self interactions for spin $0$ and spin $1$ fields, and the operator contributing to fermion-vector scattering. As we will use these operators to constrain the scale $\lnm$ that characterizes RBG models, we are interested on scattering experiments on Earth's surface. As explaiend in Sec. \ref{sec:RBG}, in this scenario the space-time metric of an RBG theory can be written as
 \begin{equation}\label{MinkowskiMetricPerturbed}
g_{\mu\nu} = \eta_{\mu\nu} +\frac{1}{\lnm^4}\lr{ \al T q_{\mu\nu}+\beta T\mn}+\cO(\lnm^{-8})\; ,
\end{equation}
which is obtained from \eqref{metricstressenergy} after neglecting Newtonian and Post-Newtonian corrections to $q_{\mu\nu}$. Up to order $\cO({\lnm^{-8}})$ corrections, its determinant is thus given by
\begin{equation}\label{detmetric}
\sqrt{-g}=1+\frac{4\al+\be}{2\lnm^4}T+\cO(\lnm^{-8}) .
\end{equation}
 Notice that, as the connection is the Levi-Civita connection of $q\mn$, and we have $q\mn\approx\eta\mn$, the connection symbols will vanish up to Newtonian and post-Newtonian corrections, which will be neglected in particle physics experiments on Earth's surface\footnote{To be more exact, if we consider fermions, they will contribute as a source of torsion to the connection, and there would arise torsion-induced interactions as explained in \cite{Kibble:1961ba,Hehl:1976kj}. However, it has been recently argued that such interactions are beyond current experimental reach \cite{Boos:2016cey}, and therefore we will neglect torsion in our discussion.}. \\

The ingredients that we need in order to construct the effective Lagrangian are equation \eqref{MinkowskiMetricPerturbed}, an expression for the connection\footnote{Notice that it vanishes up to torsion corrections generated by the fermions.}, and a matter action. We will do this for the actions that describe spin 0, 1/2 and 1 fields. Let us first start with the covariant Lagrangian for a (complex) minimally coupled scalar field in an arbitrary non-Riemannian space-time, with an arbitrary potential, and which can in principle interact with gauge bosons through its covariant derivative
\begin{equation}\label{scalaraction}
\lag_{s=0}=\sqrt{-g}\lrsq{g^{\al\be}\tilde{\na}_\al\Phi^*\tilde{\na}_\be\Phi+V_0}. 
\end{equation}
Here $\tilde{\na}$ takes into account the standard space-time covariant derivative together with the possible gauge interactions and $V_s$ corresponds to a generic potential term for a field with spin $s$, with $s=0$ in the scalar case. By making use of \eqref{metricstressenergy} and \eqref{detmetric} we can expand the above Lagrangian around a Minkowski background plus perturbations suppresed by powers of $\lnm^{-4}$. After performing such expansion the scalar Lagrangian \eqref{scalaraction} reads
\begin{equation}\label{scalarmink}
\lag_{s=0}=\eta^{\al\be}D_{\al}\Phi^*D_{\be}\Phi+V^{(0)}_0+\lag_{s=0}^Q \, , 
\end{equation}
where, given that $V_s$ can depend on the metric in the most general case, we have defined $V_s=\sum_{n=0}^\infty \lnm^{-4n}\,V_s^{(n)}$. Notice that the first two terms in \eqref{scalarmink} are the usual Lagrangian for the same complex scalar field in Minkowski space  and with the same gauge interactions (appearing inside $D_\mu$ now) and the same potential\footnote{Notice that $V_s|_{g\mn=\eta\mn}\equiv V_s^0$.} and $\lag_{s=0}^Q$ stands for a non-metricity induced contact interaction term between the scalar field and the matter stress-energy tensor which to lowest-order in $1/\lnm$ takes the form
\begin{align}\label{scalarpert}
\begin{split}
\lag_{s=0}^Q=\frac{1}{2\lnm^4}\lrsq{(2\al+\be)T\eta^{\mu\nu}-2\be T^{\mu\nu}}D_{\mu}\Phi^*D_{\nu}\Phi+\frac{1}{\lnm^4}\lrsq{\frac{4\al+\be}{2}V_0^{(0)} T+ V^{(1)}_0}+\cO(\lnm^{-8}).
\end{split}
\end{align}
Given that the interacting terms are a product of the matter stress-energy tensor and some piece of the matter Lagrangian, they respect all the symmetries of the original matter action. At the same time they describe interactions between the scalar field and all the matter fields in the model present in the stress-energy tensor. Notice that even in the free field case where there are no gauge interactions and or potential term in the original action, there arise new interactions for $\Phi$. Then, as shown already in \cite{Afonso:2018bpv}, any (minimally coupled) free spin 0 field in an RBG theory can be identified with a scalar field with the same quantum numbers that interacts with all the fields in the matter Lagrangian but evolves according to GR.\\

In order to discuss spin 1/2 fields, let us write down the Lagrangian for a minimally coupled spin $1/2$, (possibly) interacting with gauge fields and with an arbitrary potential in a general non-Riemannian space-time
\begin{equation}\label{BDlag}
\lag_{s=1/2}=\sqrt{-g}\lrsq{\frac{1}{2}\vid{a}{\mu}\lr{\bpsi\ga^a(\tilde{\na}_\mu\psi)-(\tilde{\na}_\mu\bpsi)\ga^a\psi}+V_{1/2}}.
\end{equation}
Here $\tilde{\na}_\mu\psi\equiv\lr{\pa{\mu}-\Ga_\mu-B_\mu}\psi$, where $\Ga_\mu\equiv {\omega_{\mu}}^{ab}\sigma_{ab}$ is the space-time spinor connection and $B_\mu$ represents arbitrary gauge interactions. As usual  ${\omega_{\mu}}^{ab}\equiv \frac{1}{2}(\pa{\mu}\viu{b}{\al}+\viu{b}{\be}{\Ga_{\mu\al}}^\be)\eta^{ac}\vid{c}{\al}$, $\sigma_{ab}\equiv 1/4\lrsq{\ga_b,\ga_a}$, and $\viu{a}{\mu}$ are the tetrads, here defined by $g\mn=\viu{a}{\mu}\viu{b}{\nu}\eta_{ab}$. Using \eqref{MinkowskiMetricPerturbed} and up to lowest-order in $1/\lnm$ the tetrads are given by
\begin{equation}\label{vierbeinexp}
\vid{a}{\mu}={\delta_a}^\mu-\frac{1}{2\lnm^4}\lr{\al T{\delta_a}^\mu+\be {T_a}^\mu}+\cO(\lnm^{-8}) \, .
\end{equation}
Having neglected torsion, this allows us to re-write the spinor lagrangian \eqref{BDlag} as 
\begin{equation}\label{BDlagpert}
\lag_{s=1/2}=\frac{1}{2}\lr{\bpsi\ga^\mu(D_{\mu}\psi)-(D_{\mu}\bpsi)\ga^\mu\psi}+V^{(0)}_{1/2}+\lag^{Q}_{s=1/2},
\end{equation}
where $D_\mu\psi=\lr{\pa{\mu}-B_\mu}\psi$ accounts for the gauge interactions. Notice that, again, we have the Lagrangian for the same spin 1/2 field in Minkowski space (with the same gauge interactions and potential) and a non-metricity-induced interaction Lagrangian $\lag_{s=1/2}^Q$. As in the scalar field case, this interaction term also describes a contact interaction term between the spin $1/2$ field and the matter stress-energy tensor which respects all the original symmetries of the matter sector. To lowest-order in the $1/\lnm$ expansion, this interaction term is given by
\begin{equation}\label{spinorpert}
\lag^Q_{s=1/2}=\frac{1}{4\lnm^4}\lrsq{\lr{3\al+\be}T\eta^{\mu\nu}-\be {T}^{\mu\nu}}\lrsq{\bpsi\ga_\mu(D_{\nu}\psi)-(D_{\nu}\bpsi)\ga_\mu\psi}+\frac{1}{\lnm^4}\lrsq{\frac{4\al+\be}{2}TV^{(0)}_{1/2}+V_{1/2}^{(1)}}+\cO(\lnm^{-8}).
\end{equation}
This again describes an interaction between the $\psi$ field and all matter fields through the stress-energy tensor. Notice that, as in the scalar field case, even in absence of potential and gauge interactions in the original action, the field $\psi$ becomes an interacting field. As a remark, let us point out that unlike in the scalar case, the equivalence between RBG with a free spin 1/2 field and GR with an interacting spin 1/2 field has not yet been found at a full non-perturbative level. This is due to the fact that the spin 1/2 fields couple to the affine connection. Nonetheless both the discussion on non-minimally coupled matter fields in \cite{Afonso:2017bxr}, as well as the work developed in \cite{Latorre:2017uve}, suggest that some analogy can also be found in the fermionic case. {However, as it was explained in section \ref{sec:RBG} one would have to take into account the torsion sourced by the fermions through their hypermomentum current, which will introduce corrections through the covariant derivatives in \eqref{BDlag}.  The torsion tensor sourced by the fermions will be of the form $S_{\al\mu\nu}=-i\kappa^2|\Omega|^{-1/2}\epsilon_{\al\mu\nu\sigma}\lrsq{\bar{\Psi}\gamma^\sigma\gamma_5\Psi}$. Thus, expanding $|\Omega|=1+\cO(\lnm^{-4})$ this introduces a well known (see {\it e.g.} \cite{Kibble:1961ba} 4-fermion interaction which is suppressed by the Planck scale and also $\cO(\lnm^{-8})$ corrections to the effective interaction lagrangian \eqref{spinorpert}. Since we are considering terms only up to $\cO(\lnm^{-4})$, these corrections can be neglected in this work if the scale $\lnm$ is below the Planck mass. In the case that $\lnm$ (and thus $\Lambda_{\rm RBG}$) is of the order of the Planck mass, one would need to consider the extra 4-fermion interaction induced by torsion through the fermionic hypermomentum. However, given that we will use data on Light-by-light and Compton scattering to constrain $\lnm$, and that this torsion-induced 4-fermion interaction term does not contribute to these processes at tree-level, these corrections are of sub-leading order even in the case $\lnm\sim M_{\rm P}$.}\\

The last case that we will develop is the spin $1$ field with an arbitrary potential. The corresponding Lagrangian is given by
\begin{equation}\label{VectorLag}
\lag_{s=1}=\sqrt{-g}\lrsq{\frac{1}{4}g^{\mu\nu}g^{\al\be}{F}^\dagger_{\mu\al}{F}_{\nu\be}+V_1},
\end{equation}
where ${F}_{\mu\nu}=(\text{d}A)_{\mu\nu}$. From \eqref{metricstressenergy} and \eqref{detmetric}, we can once more expand the spin 1 Lagrangian \eqref{VectorLag} around a Minkowski background, obtaining
\begin{align}\label{Vectormink}
\lag_{s=1}=\lrsq{\frac{1}{4}\eta^{\mu\nu}\eta^{\al\be}F^\dagger_{\mu\al}F_{\nu\be}+V^{(0)}_1}+\lag_{s=1}^Q.
\end{align}
 As in the two previous cases, the result of the expansion is the usual Lagrangian for the same vector field in Minkowski space-time with the same potential, and an interaction term $\lag_{s=1}^Q$ that describes an interaction between the vector field and the stress-energy tensor. This term can be written to lowest-order in inverse powers of $\lnm$ as
\begin{align}\label{vectorpert}
\begin{split}
\lag_{s=1}^Q=\frac{1}{4\lnm^4}\lrsq{(2\al+\be)T\eta^{\mu\nu}\eta^{\al\be}-2\be \eta^{\mu\nu}T^{\al\be}}F^\dagger_{\mu\al}F_{\nu\be}+\frac{1}{\lnm^4}\lrsq{\frac{4\al+\be}{2}T V_1^{(0)}+V^{(1)}_1}+\cO(\lnm^{-8}).
\end{split}
\end{align}
Once again we find that even in the free vector field case new contact interactions between the vector field and all the matter fields in the action arise.  These interactions respect all the symmetries of the original matter action by construction. The correspondence between RBG with a free spin 1 field and GR with an interacting spin 1 field has been partially established in \cite{Afonso:2018mxn}, where it is shown that, in the electrostatic case (no magnetic field), Maxwell's theory coupled to an RBG is equivalent to some non-linear electrodynamics coupled to GR. Although the discussion presented here is at a perturbative level, a recent work shows that a non-perturbative correspondence can be established even in full generality \cite{Delhom:2019zrb}.\\

We want to emphasize that the kind of analysis that we have explicitly carried out for spin $0$, $1/2$ and $1$ fields can be generalized in a straightforward way to arbitrary spin fields in the following sense: given that the metric tensor couples to fields of every spin\footnote{This implies that matter fields of any spin enter the matter stress-energy tensor.}, the non-metricity-induced corrections to the space-time metric \eqref{metricstressenergy} will induce effective interactions between any kind of matter field and the matter stress-energy tensor, regardless of their spin or other quantum numbers. As shown above, the form of the interactions can be obtained by using the $1/\lnm$ expansion of \eqref{metricstressenergy} in the corresponding Lagrangian. Then, if one neglects Newtonian and post-Newtonian corrections, it is clear that one would be able to write the covariant Lagrangian for an arbitary spin field as a sum of the Lagrangian for that same field in Minkowski space-time plus corrections. These corrections would be given by a tower of contact interaction terms induced by the non-metricicity-related corrections that appear in \eqref{metricstressenergy} and are encoded in the $\lag^Q$ term. As in the cases calculated above, these corrections will describe an effective interaction between the matter stress-energy tensor $T\mn$ and the corresponding arbitrary spin field, thus respecting all the symmetries of the original matter Lagrangians. In general, the implications of these $\lag^Q$ are the following: 1) Below the scale $\lnm$, they describe a series of perturbative interactions that can be directly related to the non-metricity tensor within RBG theories. The energy scale $\lnm$ characterizes the scale at which non-metricity becomes non-perturbative and the expansions \eqref{metricstressenergy} and \eqref{nonmetricitystressenergy} (and therefore the perturbative analysis) break down.  2) they can also be understood as departures from GR associated with non-metricity-induced corrections to the space-time metric which are sensitive to the local distribution of energy-density instead of integrated energy-density, and which are characterized by the scale $\lnm$. Notably these departures are different in nature than the differences in the post-Newtonian behaviour of RBG models, which are associated to integrated energy-density (curvature effects) rather than to the local distribution of energy-density (non-metricity effects), and which are characterized by the Planck scale instead of $\lnm$.

\section{Observational constraints to E\lowercase{i}BI and the full RBG family}\label{sec:ExpCons}

The next natural step after the above discussion is to understand the possible observability of these interactions arising in RBG models and constrain the only free parameter of a given RBG model, i.e. the non-metricity scale $\lnm$. It is beyond the scope of this paper to perform a systematic analysis of the effect of these interactions in all relevant observables, although a detailed study of the contribution of these terms will be subject of forthcoming works. Nonetheless, it is illustrative to derive some particular operators, as we can then use them to confirm the constraints already found in \cite{Latorre:2017uve} by using different data.\\

\subsection{Operators that contribute to self-scalar, self-vector and vector-fermion interactions}

Let us derive the operators corresponding to self-interactions for spin $0$ and spin $1$ from the Lagrangians \eqref{scalarpert} and \eqref{vectorpert} respectively and the vector-fermion operator from \eqref{spinorpert} and \eqref{vectorpert}. Given that the terms in which $T\mn$ appears in \eqref{scalarpert} and \eqref{vectorpert} are of order  $\cO(\lnm^{-4})$, only the Minkowskian stress-energy tensor will contribute to these operators to lowest-order in $1/\lnm$. The stress-energy tensor $T_{\mu\nu}^{(s)}$ for spin $s=\{0,1/2,1\}$ fields can be derived from \eqref{scalaraction} and \eqref{VectorLag} from the usual definition $T^{(s)}_{\mu\nu}\equiv-\frac{2}{\sqrt{-g}}\frac{\partial\lag_s}{\partial g^{\mu\nu}}$.  We can obtain the Minkowskian stress-energy tensors by performing the substitution $(\na_\mu,\,g\mn)\longrightarrow(D_{\mu},\,\eta\mn)$ in the curved space $T_{\mu\nu}$. For spin 0, 1/2 and 1 they are given by

\begin{align}
&T^{(0)}_{\mu\nu}=\eta_{\mu\nu}\lrsq{D^{\al}\Phi^*D_\al\Phi+V_0}-2\lrsq{D_{(\mu}\Phi^*D_{\nu)}\Phi+\bar{\frac{\partial V_0}{\partial g^{\mu\nu}}}}+\cO(\lnm^{-4}) \, , \label{scalartmunu}\\
&T^{(1/2)}_{\mu\nu}=\eta_{\mu\nu}\lrsq{\frac{1}{2}\lr{\bpsi \ga^\al (D_{\al}\psi)-(D_{\al}\bpsi)\ga^\al\psi}+V_{1/2}}-\lrsq{\bpsi\ga_{(\mu}(D_{\nu)}\psi)-(D_{(\nu}\bpsi)\ga_{\mu)}\psi+2\bar{\frac{\partial V_{1/2}}{\partial g^{\mu\nu}}}} +\cO(\lnm^{-4})  \, , \label{spinortmunu}\\
&T^{(1)}_{\mu\nu}=\eta\mn\lrsq{\frac{1}{4}F^\dagger_{\al\be}F^{\al\be}+V_1}-\lrsq{F^\dagger_{(\mu|\al|}{F_{\nu)}}^\al+2\bar{\frac{\partial V_{1}}{\partial g^{\mu\nu}}}}+\cO(\lnm^{-4}).  \label{vectortmunu}
\end{align}
Here, the bar over the terms $\partial \bar{V_s}/\partial g$ indicates that the replacement $(\na_\mu,\,g\mn)\longrightarrow(D_{\mu},\,\eta\mn)$ has to be  made after taking the functional derivative. The corresponding traces in 4 dimensions are:
\begin{align}
&T^{(0)}=2D^{\al}\Phi^*D_\al\Phi+4V_0-2\eta^{\mu\nu}\bar{\frac{\partial V_0}{\partial g^{\mu\nu}}} +\cO(\lnm^{-4})     \label{scalarttrace} ,\\
&T^{(1/2)}=\lrsq{\bpsi \ga^\al (D_{\al}\psi)-(D_{\al}\bpsi)\ga^\al\psi}+4V_{1/2}-2\eta^{\mu\nu}\bar{\frac{\partial V_{1/2}}{\partial g^{\mu\nu}}}   \label{spinorttrace} ,\\
&T^{(1)}=4V_1-2\eta^{\mu\nu}\bar{\frac{\partial V_{1}}{\partial g^{\mu\nu}}}+\cO(\lnm^{-4}).\label{vectortrace}
\end{align}

After substitution of the stress-energy tensor and the trace of a field of a given spin in the Lagrangian of any other field, new interactions between them will arise. Let us exemplify this with the following cases: 

\begin{itemize}
    \item[1)] Effective self interactions for a Higgs-like scalar field, i.e. with a potential not dependent of the metric $\partial V_0/\partial g~=~0$. 
    \item[2)]Effective self interactions for a photon-like field, i.e. a $U(1)$ vector field with $V_1=0$.
    \item[3)] Effective interactions between a photon-like field and a massless fermion, i.e. a fermion with $V_{1/2}=0$ but otherwise arbitrary.
\end{itemize}

The corresponding operators for scalar self-interactions are obtained after substituting the stress-energy tensor for a complex scalar field \eqref{scalartmunu} and its trace \eqref{scalarttrace} in the effective Lagrangian for the same complex scalar field \eqref{scalarpert}. Then, setting  $\partial V_0/\partial g=0$, for case 1) we have
\begin{equation}\label{selfscalar}
\begin{split}
\lag_{\Phi\Phi}^Q=\frac{1}{\lnm^{4}}\bigg[&(6\al+\be)(D_{\mu}\Phi^*D^{\mu}\Phi)(D_{\nu}\Phi^*D^{\nu}\Phi)-\beta (D_{\mu}\Phi^*D^{\mu}\Phi^*)(D_{\nu}\Phi D^{\nu}\Phi)\\
&+(21\al+5\be)V_0 D_{\mu}\Phi^*D^{\mu}\Phi+(8\al+\be)(V_0)^2\bigg].
\end{split}
\end{equation}
Notice that if the scalar field potential $V_0$ does not depend on the metric, we have $V_0^{(0)}\equiv V_0$, as all the $n>0$ terms in the expansion $V_0=\sum_{n=0}^\infty \lnm^{-4n}V_0^{(n)}$ vanish identically. For the vector self-interactions case, we must now substitute the stress-energy tensor for a $U(1)$ field and its trace in the effective Lagrangian for that same vector field. Setting  $V_1=0$ we get the Lagrangian
\begin{equation}\label{4VLag}
\lag_{AA}^Q = -\frac{\be}{8\lnm^4}\lrsq{F_{\mu\nu}F^{\mu\nu}F_{\al\be}F^{\al\be}-4F_{\mu\nu}F^{\nu\al}{F^\mu}_{\sig}{F^\sig}_{\al}},
\end{equation}
which is a particular of the well known $C$, $P$, Lorentz and gauge invariant effective Lagrangian describing photon-photon collisions below the mass scale of some charged fermion. Notice that while the Euler-Heisenberg Lagrangian \cite{Pich:1998xt,Euler:1936oxn,Heisenberg:1935qt} obtained by integrating out a massive lepton in the QED path integral gives a relation $b/a=-14/5$ , the above Lagrangian satisfies $b/a=-4$. The operators above contribute, for instance, to the process $\gamma\ga \to \ga\ga$ at tree level. This contribution will be later used to set a lower bound on $\lnm$. Finally, for case 3), we have to substitute the stress-energy tensor of the photon in the effective Lagrangian of a spin 1/2 field \eqref{spinorpert} (with $V_{1/2}=0$ and coupled or not to the photon through $D_\mu$), and then do the same with the stress-energy tensor of that same spin 1/2 field in the effective Lagrangian for the photon \eqref{vectorpert}. We then get an effective dimension-8 operator describing an interaction between the fermion pair and a photon pair of the form

\begin{equation}\label{vector-fermion}
\begin{split}
\lag^Q_{\psi A}=-\frac{9\be}{4\lnm^4}F^{\mu\al}{F^\nu}_\al\lrsq{\bpsi\ga_\mu(D_{\nu}\psi)-(D_\nu\psi)\ga_\mu\psi}+\cO\lr{\lnm^{-8}}.
\end{split}
\end{equation}
This operator is generic for every fermion whether it has a non-vanishing electric charge or not. Therefore it generates, for instance, an effective photon-neutrino coupling at tree level, or corrections to matrix elements for other processes that are already tree level in the SM like for example Compton scattering. An effective interaction between neutrinos and photons could have wide implications. To name one, it could change the relation between the Cosmic Microwave Background and the Cosmic Neutrino Background temperatures, which could also be used to constrain $\lnm$.\\

\subsection{Experimental constraints to $|\beta|^{-1/4}\lnm$ in generic RBGs}

Even though almost every process is sensitive to contributions appearing in Ricci-based gravity theories,  obtaining constraints for the scale $\Lambda_Q$ is not straightforward. Corrections induced in the vertices and in the partition distributions functions of gluons and quarks make it very difficult to study processes in which particles are produced via $p\bar{p}$ production. This makes high-energy data from LHC not convenient for this study and requires to consider experimental bounds at lower energies. Thus we will use for that purpose current data on light-by-light and Compton scattering. \\

Light-by-light scattering occurs at loop level in the SM and therefore 
it is very suppressed \cite{Fichet:2014uka,Witten:1977ju,Terazawa:1973tb}. 
Therefore, it could be interesting to obtain lower-energy bounds from experiments searching photon self-interactions. This has been done with X-ray pulses \cite{Inada:2014srv} obtaining an upper bound for the cross section which can be used to constrain $\lnm$. The differential and total cross sections for $\ga\ga \to \ga\ga$ that one obtains from the RBG corrections \eqref{4VLag} at tree level are given by
\begin{eqnarray}
\frac{d\sig^Q_{\gamma\ga\rightarrow\ga\ga}}{d\Omega} &=& \lr{\frac{\be}{8 \lnm^4}}^2\frac{1}{256\pi^2}s^3\lrsq{512+32\lr{\lr{1-\cos{\theta}}^4+ \lr{1+\cos{\theta}}^4}} \label{photonphotondiffcrossec} , \\
\sig^Q_{\gamma\ga\rightarrow\ga\ga} &=&\lr{\frac{\be}{8 \lnm^4}}^2\frac{56}{5\pi}s^3, \label{photonphotoncrossec}
\end{eqnarray}
to lowest-order in $1/\lnm$. By demanding \eqref{photonphotoncrossec} to be in agreement with the current experimental limit of $\gamma\ga \to \ga\ga$ at 6.5 keV, $\sig^{exp}_{\gamma\ga \to \ga\ga} < 1.9 \times 10^{-27}$ m$^2 $ \cite{Yamaji:2016xws}, we can set a lower bound
\bee\label{Lightbylightbound}
 |\be|^{-1/4} \lnm > 23.3 \; \text{keV}.
\ee

\begin{figure}
\centering
\includegraphics[scale=0.8]{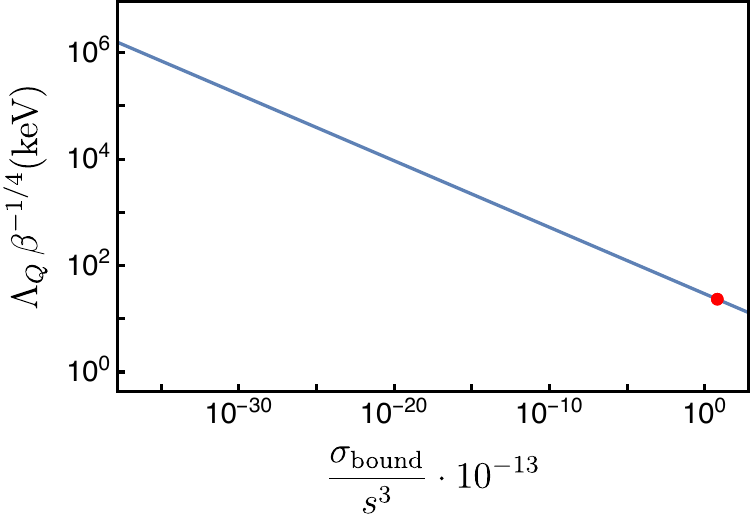}
\caption{Expected bounds on $\Lambda_Q$ for different values of the ratio $\frac{\sigma_{\text{bound}}}{s^3}$ in logarithmic scale. Our bound is denoted by a red point.}
\label{fig:photonsprospects}
\end{figure}

Notice that the value of $\beta$ is specified once a specific RBG model is chosen, allowing to constrain directly the energy scale $\lnm$. Due to the difference in energies at which Bhabha and photon-photon scattering are currently tested, and the unobservability of photon self interactions in the keV range with current experimental precision, the bound obtained in \eqref{Lightbylightbound} is considerably weaker than the one obtained from electron-positron scattering in \cite{Latorre:2017uve}. 
However, future experiments searching for light-by-light scattering in the keV range could help to tighten current constraints in RBG models, provided that a substantial increase in the experimental resolution is achieved. If precision is not improved, higher-energy experiments will allow us to obtain stringent bounds to RBG. In Fig.~\ref{fig:photonsprospects} we can see how the limit would change if the precision is improved or the energy scale changes. For instance, keeping the same upper limit $\sigma_{\text{bound}}$ while increasing the energy scale of the experiment in an order of magnitude, bounds will improve roughly in one order of magnitude. Actually, light-by-light scattering has been measured by ATLAS at $\mathcal{O}$(GeV) for the invariant mass in LHC Pb-Pb collisions \cite{Aaboud:2017bwk}. After an involved analysis, these data allow to set a lower bound to the mass scale of Born-Infeld electrodynamics through its lowest order corrections to Maxwell electrodynamics \cite{Ellis:2017edi}. In the same spirit, they would allow to set bounds to $\beta/\lnm^4$ through our Lagrangian \eqref{4VLag}. Although performing an in-depth analysis of LHC data is out of the scope of this work, a rough order-of-magnitude estimate allows us to translate the bounds in \cite{Ellis:2017edi} to bounds on generic RBGs, obtaining an approximate bound of $ |\be|^{-1/4} \lnm  \gtrsim 140 \; \text{GeV}$, which is in agreement with the bounds from Bhabha scattering obtained in \cite{Latorre:2017uve}.

\begin{figure}
    \centering
    \includegraphics[scale=1.]{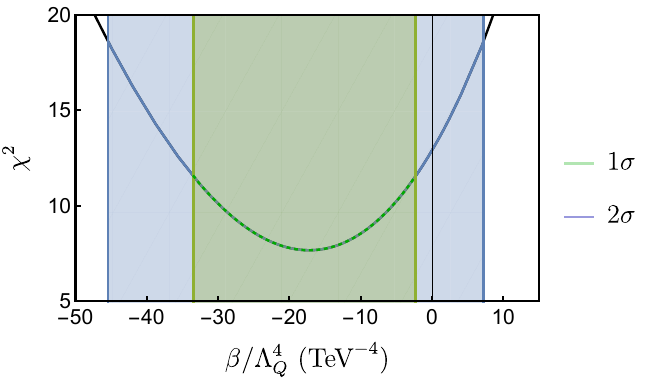}
    \caption{ Value of the $\chi^2$ for different values of $\beta/\Lambda_Q^4$. The green and blue bands indicate the allowed values of $\beta/\Lambda_Q^4$ at $1\sigma$  and $2\sigma$ probability respectively.}
    \label{fig:chi_Compton}
\end{figure}

We can however look for a tighter bound in more clean high-energy experiments. To that end, let us consider Compton scattering as a probe for RBG corrections to the vector-fermion cross-section. The most recent data for the cross-section of Compton scattering comes from the L3 collaboration \cite{Achard:2005cs}, where the process was measured at different energies as  in Tab.~\ref{tab:expcompt}. The differential 
for Compton scattering in RBG theories can be obtained from the SM Lagrangian together with \eqref{vector-fermion}, and is given by
\begin{eqnarray}
\nonumber\frac{d \sigma^{Q}_{e^{-} \gamma \to e^{-} \gamma}}{d \Omega} &=& \frac{1}{256\pi^2 s} \left[\frac{81}{16}
\left(\frac{\beta}{\Lambda_Q^4}\right)^2(\cos\theta+1)(\cos^2\theta+2 \cos\theta+5)s^4  \right. \\
&+& \left.\frac{9}{2}\left(\frac{\beta}{\Lambda_Q^4}\right)\left(3\cos^2\theta+2\cos\theta+11\right)Q_e^2 s^2 + 4 Q_e^4 \frac{\cos^2\theta + 2\cos\theta + 5}{\cos\theta +1}  \right] .
\label{eq:comptondiff}
\end{eqnarray}
where $Q_e$ is the charge of the electron. Note that in Tab.~\ref{tab:expcompt}, only the region of the phase space in which $|\cos\theta|<0.8$ is considered. This will be taken into account when placing the bounds on $\lnm$. As in Tab.~\ref{tab:expcompt} we have the experimental value measured at 12 different energies, it is convenient to combine all these measurements performing a $\chi^2$ test with 10 degrees of freedom. 
As Fig.~\ref{fig:chi_Compton} shows, by studying the probability of the resulting $\chi^2$ function we can exclude the quantity $\beta / \Lambda_Q^4$ up to a certain probability.  The green  
and blue bands contain the $1\sigma$ and $2 \sigma$ probability respectively.\\

\begin{table}[h]
\setlength{\tabcolsep}{20pt}
\centering
\begin{tabular}{c c c}
\hline
$\sqrt{s}$ (GeV) & $\sigma_{e^-\gamma \rightarrow e^-\gamma}^{exp}$  (GeV$^{-2}$) & $\sigma_{e^-\gamma \rightarrow e^-\gamma}^{QED}$ (GeV$^{-2}$)\\
\hline 
21 & 771.2$\pm$21.6 & 764.8  \\ 
29.8 & 370.6$\pm$11.3 & 381.1  \\
39.7 & 213.2$\pm$5.4 & 214.7  \\
49.7 & 128.7$\pm$3.9 & 136.7 \\
59.8 & 95.0$\pm$3.5 & 94.6  \\
69.8 & 70.6$\pm$2.9 & 69.4  \\
79.8 & 55.2$\pm$2.6 & 53.1  \\
92.2 & 38.8$\pm$2.2 & 39.8  \\
107.2 & 27.3$\pm$2.2 & 29.4  \\
122.3 & 20.0$\pm$2.1 & 22.6  \\
137.3 & 17.3$\pm$2.1 & 17.9  \\
159.3 & 9.1$\pm$2.0 & 13.3  \\
\hline
\end{tabular}
\caption{Experimental values and SM prediction of the cross section for Compton scattering taking from \cite{Achard:2005cs}.} 
\label{tab:expcompt} 
\end{table}

In Fig.~\ref{fig:chi_Compton} the full $1\sigma$ probability is in the region $\beta < 0$. While the implications of the sign of this parameter are not clear in general, its role in some relevant models is well understood in cosmological and black hole scenarios. At $2\sigma$  we get different results for different signs of $\beta$:
\begin{align}
 & |\beta|^{-1/4} \Lambda_Q   > 0.39 \text{ TeV}  \qquad \text{with } \beta < 0,
\label{2scomptonneg}\\
 & |\beta|^{-1/4} \Lambda_Q  > 0.61 \text{ TeV}  \qquad \text{with } \beta > 0 .
\label{eq:2scomptonpos}
\end{align}

These are bounds for a general RBG theory. Once a specific RBG model is chosen, the value of $\beta$ is set and the bound is translated to the only free parameter of the theory, which is always related to the non-metricity scale $\lnm$. Note that in this case the SM,  corresponding to $\beta/\Lambda_Q^4 = 0$ is already in the $2\sigma$ probability region. That means that at $1\sigma$ the values of  $\beta/\Lambda_Q^4$ giving a lower value of the $\chi^2$ (higher probability) will be negative compensating the SM contribution. As mentioned before, at $2\sigma$ the SM is already in agreement with the data so positive values of $\beta$ give bounds in this region.

\subsection{Constraints to Eddington-inspired Born-Infeld gravity}

As a particular example, let us consider a widely discussed RBG model named Eddington-inspired Born-Infeld theories (EiBI, see \cite{BeltranJimenez:2017doy} for a recent review), and in units $G=c=1$, it is defined by the action
\begin{equation}
S_{\text{EiBI}}=\pm\frac{2}{\kappa}\int\text{d}^4x\lrsq{\lr{\sqrt{-\det\lr{g_{\mu\nu}\pm\kappa R_{(\mu\nu)}}}}-\lambda\sqrt{-\det(g_{\mu\nu})}}.
\end{equation}
where $\kappa$ is a common parametrization of $\lnm$ in the astrophysics and cosmology literature and it is related to the non-metricity scale as $\kappa=2c^7\hbar^3\lnm^{-4}$. The values of $\beta$ within EiBI deppend on the sign in front of the $\kappa$ parameter, which has the values $\beta=1$ for the minus sign and $\beta=-1$ for the plus sign. In EiBI, while $\beta=1$ leads to a bouncing cosmology, $\beta=-1$ describes a cosmology in which an asymptotically Minkowski past region connects with the present contracting branch \cite{Banados:2010ix,BeltranJimenez:2017uwv}.  Interestingly, both solutions avoid the Big Bang singularity\footnote{Nonetheless, a potential Big Rip singularity could arise if phantom dark energy is considered within EiBI \cite{Bouhmadi-Lopez:2013lha}.}, although as
found in \cite{BeltranJimenez:2017uwv}, the propagation of gravitational waves (GWs) generally presents instabilities in these cosmological models. In particular, Beltran \textit{et.al.} show that for $\beta>0$ GWs develop instabilities at the bounce due to the fact that the propagation speed diverges and the friction term vanishes, signaling a strong coupling problem. On the other hand, for the asymptotically Minkowski solution (where $\beta<0$), they show that the pathologies are due to the vanishing of the propagation speed, which could in principle be avoided by including higher derivative terms. Regarding spherically symetric solutions, while $\beta=-1$ are generally singular \cite{Banados:2010ix,Olmo:2013gqa}, the $\beta=1$ branch remarkably admits non-singular wormhole space-times when coupled to Maxwell electrodynamics.\\

 The above bounds for a general RBG model can be easily translated to the EiBI theory, where $\beta=\pm1$, finding
 \begin{align}
 &\Lambda^{BI}_Q  > 0.39 \text{ TeV}  \qquad \text{with } \kappa > 0,
\label{2scomptonneg}\\
 &\Lambda^{BI}_Q  > 0.61 \text{ TeV}  \qquad \text{with } \kappa < 0 ,
\label{eq:2scomptonpos}
\end{align}
which in order to make contact with the astrophysics and cosmology literature can be translated into 
\begin{align}
    &|\kappa|<3.5\times10^{-14}\text{m}^5 \text{kg}^{-1}\text{s}^{-2}.\\
    &|\kappa|<5.5\times 10^{-15}\text{m}^5 \text{kg}^{-1}\text{s}^{-2}.
\end{align}
 The bound obtained here is of the same order as the one obtained in \cite{Latorre:2017uve}, and improves in 6 orders of magnitude the bound for the scale $\lnm$ (or 12 orders of magnitude for the $\kappa$ parameter) other constraints obtained from astrophysical or nuclear physics \cite{Avelino:2012ge,Avelino:2012qe,Avelino:2019esh} phenomena.

\section{Outlook}

In this paper we have generalized the results presented in \cite{Latorre:2017uve} for spin 1/2 fields to spin 0 and spin 1 fields, and explained why a generalization to arbitrary spin is straightforward. Concretely, we have shown that minimally coupled matter fields, when coupled to an RBG gravity, develop non-trivial interactions. These interactions arise even in the usual weak gravitational field limit (i.e. around a Minkowskian background), and are due to the higher-order curvature terms in the RBG Lagrangian. In this regard, notice that a recent technique has been developed \cite{Afonso:2018bpv,Afonso:2018hyj,Afonso:2018mxn} showing how minimally coupled matter fields coupled to an RBG can be mapped into GR coupled to non-linearly interacting matter fields. It is then said that RBGs admit an Einstein frame representation in which matter fields become non-linear. Thus we point out that our methods could be a perturbative version of this mapping between RBGs and GR in which the non-linearities induced by the mapping in the matter sector appear as effective interaction terms below the non-metricity scale $\lnm$. These interactions appear independently of the spin of the fields, as can be seen from the expansion \eqref{metricstressenergy} together with the fact that all matter fields couple to the metric, regardless of their spin. We have computed the effective Lagrangians up to order $\mathcal{O}(\lnm^{-8})$ corresponding to a scalar \eqref{selfscalar}, vectorial \eqref{4VLag}, and a spinor+vector \eqref{vector-fermion} matter sector. Finally we have used the vector self-interaction and the vector-fermion interaction Lagrangians to set bounds on $\lnm$ with data from Compton and photon-photon scattering. The bound for $\lnm$ obtained from $\gamma\gamma\rightarrow\gamma\gamma$ is very low due to the fact that the data are taken in the keV range, and the cross-section increases as $s^3$, nonetheless, since Compton scattering data from LEP are taken around the 100 GeV range, we obtain bounds for $\lnm$ which are of the same order as those obtained in \cite{Latorre:2017uve}, currently the most stringent bounds for RBG theories and EiBI in particular.\\

The existence of these effective interactions within RBG theories suggests to further explore the available data on particle physics experiments in order to tighten the constraints on the different RBG models that are currently under study in the modified gravity community. Furthermore, it would be interesting to find out whether these effective interactions are a particular feature of RBGs, or instead they also arise in more general gravity models, and could also be used to constrain them. Regarding this, notice that these constraints arise from the perturbative effects of the non-metricity-induced corrections to GR that arise in generic RBGs. However, full non-perturbative solutions of some particular RBG models are known \cite{Olmo:2012nx,Menchon:2017qed,BeltranJimenez:2017uwv,Afonso:2019fzv}, and it would also be of interest to understand what is the non-perturbative counterpart of these non-metricity-induced interactions that arise in the matter fields from the perspective of the Einstein frame.\\

{\bf Aknowledgements} We would like to thank to Gonzalo J. Olmo, Antonio Pich and Diego Rubiera-Garcia for their feedback and help in the elaboration of this work.

This work is supported by the Spanish project FIS2017-84440-C2-1-P (MINECO/FEDER, EU), the project H2020-MSCA-RISE-2017 Grant FunFiCO-777740, the project SEJI/2017/042 (Generalitat Valenciana), the Consolider Program CPANPHY-1205388, and the Severo Ochoa grant SEV-2014-0398 (Spain).

The work of VM and AP has been supported  by the Spanish Government and ERDF funds from the EU Commission [Grant FPA2017-84445-P], the Generalitat Valenciana [Grant Prometeo/2017/053], the Spanish Centro de Excelencia Severo Ochoa Programme [Grant SEV-2014-0398] and the DFG cluster of excellence Origin and Structure of the Universe. 

The work of AD, VM and AP is funded by Ministerio de Ciencia, Innovaci\'on y Universidades, Spain [Grant FPU15/05406],  [Grant FPU16/01911] and [Grant FPU15/05103] respectively.

\end{document}